Resonant Nernst effect in the metallic and field-induced spin density wave states of $(TMTSF)_2ClO_4$

E. S. Choi[a], J. S. Brooks[a], H. Kang[b], Y. J. Jo[b], W. Kang[b]

[a]NHMFL/Physics, Florida State University, Tallahassee, FL 32310, USA

[b]Dept. of Physics, Ewha Womans University, Seoul 120-750, Korea

Abstract

We examine an unusual phenomenon where, in tilted magnetic fields near magic angles parallel to crystallographic planes, a "giant" resonant Nernst signal has been observed by Wu et al.[Phys. Rev. Lett. **91** 56601(2003)] in the metallic state of an organic conducting Bechgaard salt. We show that this effect appears to be a general feature of these materials, and is also present in the field induced spin density wave phase with even larger amplitude. Our results place new restrictions on models that treat the metallic state as an unconventional density wave or as a state with finite Cooper pairing.

The Bechgaard salts, (TMTSF)$_2$X (X=PF$_6$, ClO$_4$, AsF$_6$, ReO$_4$,…) continue to attract attention due to the proximity of novel ground states including antiferromagnetic spin density waves (SDW), unconventional superconducting (SC) and metallic states, and magnetic field induced spin density wave (FISDW) states [1,2]. Recently, arguments have been made for the coexistence of the SDW, SC, and metallic states along phase boundaries [3-5], for the possibility that the metallic state is an unconventional SDW phase [6], and even for the survival of p-wave Cooper pairing in the metallic state [7].

Most studied are (TMTSF)$_2$PF$_6$ (under pressure to remove the ambient pressure SDW phase) and (TMTSF)$_2$ClO$_4$. These materials are quasi-one-dimensional (Q1D) organic conductors with transfer integrals of order 250, 25, and 0.25 meV for the *a*, *b*, and *c* axis directions respectively. In the SC state, evidence for p-wave pairing exists since the critical field H$_{c2}$ exceeds the Pauli limit when the field is aligned parallel to the molecular planes [8,9], and additionally from Knight shift studies in (TMTSF)$_2$PF$_6$ [10]. Above H$_{c2}$, in the metallic state, magic angle effects [11] were first observed in (TMTSF)$_2$ClO$_4$ [12,13] (and later in (TMTSF)$_2$PF$_6$ [14]) as dips at magic angles in angular dependent magnetoresistance (ADMR) data for tilted magnetic field in the *bc*-planes, perpendicular to the most conducting *a*-axis. The magic angles occur when tanθ ≈ 1.1 *p/q*, where θ is the angle between the field and *c*-axis, and *p*, *q* are integer values of the *c* and *b* unit cell parameters respectively [12,13]. Here the field is aligned along crystal planes composed of the *a*-axis and integer combinations of the *b* and *c* unit cell parameters. Above a threshold field H$_{th}$, the FISDW state (with a cascade of subphases) is stabilized due to the quantized nesting of the Q1D Fermi surface [1,15]. Unlike the tan(θ) dependence, the threshold and subphases in the FISDW phase are driven by the field component perpendicular to the conducting *ab* planes, i.e. the "cosine law" H$_\perp$=H cos(θ).

In this Letter we examine the longitudinal (Seebeck or TEP) and transverse (Nernst) thermoelectric effects in the metallic and FISDW phases in (TMTSF)$_2$ClO$_4$ (hereafter ClO$_4$). The Nernst effect, an analogue of the Hall effect, is the transverse voltage induced by a temperature gradient in a magnetic field. It has been studied in cuprates due to its sensitivity to the vortex state, where large Nernst signal can be induced by phase slip [16,17]. On the other hand, quasi particles in normal Fermi liquids with one type of charge carriers are known to generate very small Nernst effect because of so called the "Sondheimer cancellation" [18,19]. When more than two types of carriers are involved, the cancellation is not perfect and the Nernst effect can be finite with an order of $(e/k_B)(T/T_F)(\omega_c\tau)$ where $k_B T_F$ is the Fermi energy, $\omega_c$ is the cyclotron frequency and $\tau$ is the scattering time, which is few nV/K for the Bechgaard salts [20]. Recently a resonant-like Nernst effect in $(TMTSF)_2PF_6$ (hereafter $PF_6$) has been reported by Wu et al. [20]. Here large negative dips and positive peaks are observed in a narrow range ($\pm 5°$) as the field angle is swept through magic angles. The effect was originally attributed to an electrical current "lock-in" at the magic angle where metallic behavior appears, and to non-metallic behavior away from the magic angle [20]. Currently, a model has been proposed by Ong et al.[7] involving notion of finite Cooper pairing in the metallic phase. Here the symmetry of pairs of fluctuating pancake vortices in the crystallographic planes will be broken by induced vortex density for finite in-plane field away from the magic angle directions. This can account for the magnitude and resonant nature (change of sign) of the Nernst signals even though the cross product of the temperature gradient with the field direction has not changed sign [7]. *We find that $ClO_4$ exhibits a resonant Nernst effect not only in the metallic state, but also at higher fields in the FISDW state, with a greatly enhanced magnitude.* As will be discussed below, these new findings place restrictions on models proposed to explain the anomalous Nernst signal in these materials.

Superconductivity appears at ambient pressure in $ClO_4$ below ~ 1 K when cooled slowly (10~15 mK/min in this work) through the anion ordering temperature $T_{AO}$~ 24 K. Here the tetrahedral $ClO_4$ anions order, doubling the unit cell along the *b*-axis. Although no AO occurs in $PF_6$, pressures above 6 kbar are necessary in $PF_6$ to stabilize the metallic and superconducting ground states. Under these respective conditions, both materials exhibit a metal-to-FISDW transition above a threshold magnetic field ($H_{th}$ ~ 4 T for $ClO_4$ and $H_{th}$ ~ 7 T for $PF_6$ at 10 kbar) for B//*c*. The contact configuration for the TEP and Nernst effect measurement for $ClO_4$ is shown in the inset of Fig. 1. The sample is placed in an evacuated holder sealed at room temperature. Our method [21] is different from that of Wu et al. [20] since the sample is in vacuum and the Nernst signal is measured along the $b$-axis with $\Delta T \parallel a$-axis. The sample is rotated around the $a$-axis so that the magnetic field (either forward or reversed) is in the $bc$-plane. The TEP signal is measured from the ratio of the longitudinal thermoelectric voltage drop to the temperature gradient ($\Delta V_1/\Delta T$) while the Nernst signal from the transverse voltage drop ($\Delta V_2/\Delta T$). The final TEP and Nernst data are obtained by averaging the signals from the positive and negative field runs assuming that the TEP (Nernst effect) is an even (odd) function of the magnetic field. The ADMR and Hall effect are measured on the same sample by using the sample contacts for voltage and current. The angular position of magic angle effects for ClO$_4$ in $R_{xx}$ and $R_{xy}$ agree with previous reports [12].

Fig. 1 shows the angle dependence of the Nernst signal ($N$) at different fields at 0.4 K. The down arrows indicate the FISDW threshold angle ($\theta_{th}$) determined by $H_\perp = H \cos\theta_{th} = H_{th}$, where $H_{th}$ (about 3.9 T at 0.4 K) is the threshold field. Resonant Nernst signals are observed in both the metallic ($\theta > \theta_{th}$) and FISDW ($\theta < \theta_{th}$) states. The Nernst signal increases as the sample undergoes a phase transition between the metallic and FISDW states. In the FISDW state, additional structure related to the cascade of FISDW sub-phases appears as $H_\perp$ increases for $\theta \rightarrow 0$. In contrast to the Nernst effect, the low temperature TEP signal does not show noticeable changes at magic angles in the metallic state as shown in the inset of Fig. 1.

In the metallic phase, the positions of dips at -52° ($\theta_{m-}^{(1)}$) and peaks at -43° ($\theta_{m+}^{(1)}$), in Fig. 1 are field independent. Furthermore, the mid-angle between two extremes, which is -48° ($\theta_m^{(1)}$), is the magic angle for $p=q=1$ where the field is along the $2b+c$ direction (consistent with anion ordering). Hence the resonant-like behavior with sign changes of the Nernst effect around the magic angle for ClO$_4$ is similar to that of PF$_6$ at 10 kbar [20]. For the magic angle $\theta_m^{(0)}$ for B//$c$* the Nernst signal appears to be resonant (as it is in PF$_6$ where it is accessible below $H_{th}$), but due to the lower $H_{th}$ the additional complications of the FISDW subphases, make systematic analysis difficult.

The peak-to-peak values of the Nernst signal ($N_{pp}$) around the magic angles in the metallic state were measured vs. field at 0.4 K and vs. temperature at 5 T and 5.5 T. The normalized peak-to-peak Nernst data ($N_{pp}^N = N_{pp}/H$) is shown in Fig. 2. For the values around $\theta_m^{(1)}$, $N_{pp}^N$ grows quickly for fields higher than 4.5 T and temperatures lower than 0.8K. The increasing $N_{pp}^N$ indicates that $N_{pp}$ is also non-linear function of magnetic field. The maximum $N_{pp}$ in ClO$_4$ is about 4 µV/K at 5.2 T and 0.4 K (for the first magic angle $\theta_m^{(1)}$), which is of same order of magnitude as PF$_6$ even though PF$_6$ was measured along the *a*-axis with $\Delta T \parallel$ *b*-axis [20].

When we discovered that the resonant Nernst signal in the FISDW resembles that of the metallic state, we pursued this finding at higher fields where the FISDW threshold angle ($\theta_{th}$) progressively increases with respect to the magic angles. The angular dependences of the Nernst effect at selected fields are presented in Fig. 3, where in the inset, a summary of all data clearly show the equivalence of the resonant Nernst periods for both metallic and FISDW phases. The solid vertical lines in the main panel of Fig. 3 correspond to the angles where the ADMR (measured separately) shows dips, and the dashed lines indicate the peak and dip positions for the resonant Nernst signal in the metallic phase. The second ($p/q = 2$) and the possibly the third ($p/q = 3$) magic angle effects are evident. The up arrows indicate $\theta_{th}$ and the down arrows show the angle below which the Nernst signal is angle independent. The down arrows follow $H_\perp = H \cos\theta$, where $H_\perp = 6.5 \sim 6.9$ T. Additional structures due to the FISDW subphases appear in the resonant behavior.

In Fig. 4 the field dependence of the Nernst effect is shown for fixed angles $\theta = 0$, $\theta_{m+}^{(1)}$, $\theta_m^{(1)}$ and $\theta_{m-}^{(1)}$ vs. $H_\perp = H \cos\theta$. For $\theta = \theta_{m+}$ and $\theta_{m-}$, the Nernst effect starts to deviate from the low field behavior at fields ($H_\perp = 2.3$ T for $\theta = \theta_{m-}$ and 3.1 T for $\theta = \theta_{m+}$) in the metallic state. In the FISDW state above $H_{th}$ the Nernst signal changes sign for small variations around the magic angle, and its magnitude at $\theta = \theta_{m-}$ and $\theta_{m+}$ is even larger than $\theta = 0°$. As in Fig. 3, at higher fields above $H_\perp = 6.5 \sim 6.9$ T, the Nernst effect is small for any field direction. Although perhaps coincidental, this field is in the range where the Hall resistance makes its last sign reversal above the Ribault anomaly [22], as shown in the inset of Fig. 4. The high field resonant Nernst signal for H||$c^*$ is also suppressed in PF$_6$, when the sample is in the FISDW state (0.2K, H$_\perp$=7.5 T) [20].

In the metallic phase, ClO$_4$ and PF$_6$ show a similar resonant Nernst behavior around magic angles, even with anion ordering in the case of ClO$_4$, and even though the present and previous studies were for the off-diagonal and diagonal configurations respectively. This includes similarities in the amplitude and the field and temperature dependence of N$_{pp}$. Since our measurements involve the traditional off-diagonal Nernst configuration, the resonant Nernst signal does indeed contradict the prediction of the unconventional density wave theory [6] where only peaks are predicted at magic angles. However, torque studies in ClO$_4$ [12] clearly show a peak at the magic angles. This is inconsistent with the vortex model: since the vortex magnetization **M**$_{vortex}$ ~ **H** sin($\theta_m - \theta$) changes sign going through the magic angles, the torque magnetization **M**$_{vortex}$×**H** should also change sign.

Our observation that the resonant Nernst effect persists into the FISDW phase, with larger amplitude, would indicate that either the Cooper pairing and vortex state also survive the FISDW transition, and/or that some other mechanism involving the Fermi surface nesting is involved. In addition to the high T$_c$ cuprates [17], there are reports of large Nernst effects in another organic charge transfer salt α-(BEDT-TTF)$_2$KHg(SCN)$_4$ [23,24], and in inorganic materials such as NbSe$_2$ [25] and CeCoIn$_5$ [26], all well removed from superconducting phases. Both α-(BEDT-TTF)$_2$KHg(SCN)$_4$ and NbSe$_2$ undergo a density wave transition at 8 K and 32 K respectively. In these materials the large Nernst effects were attributed to the quasiparticles, rather than vortices, in the density wave states (for α-(BEDT-TTF)$_2$KHg(SCN)$_4$ and NbSe$_2$) or near the quantum critical point (for CeCoIn$_5$).

The presence of Nernst effect in the FISDW state can be modeled by considering two parallel conduction channels, one from the nested gapped Fermi surface and the other from electron or hole carriers responsible for the Landau quantization [27]. The former carriers should be activated across the gap, so the resistance and thermoelectric tensor will be divergent with decreasing temperature, while the latter carriers may retain the character of the metallic state. The total Nernst effect will be $N_{total} = (\sigma_{FISDW} N_{FISDW} + \sigma_{Metal} N_{Metal})/(\sigma_{FISDW} + \sigma_{Metal})$, where $\sigma_{FISDW}$ ($\sigma_{Metal}$) and $N_{FISDW}$($N_{Metal}$) represents conductivity and Nernst signal in the FISDW (metallic) state respectively. Since the contribution to the total Nernst effect of the FISDW carriers is the weighted by their conductivity, the behavior of the metallic carriers may survive in the FISDW state. Although this model does not explain the enhancement and the sign reversal in the FISDW state seen in the inset of Fig. 3, the suppression of the Nernst signal at higher fields may be explained if one assumes $\sigma_{FISDW}$ becomes very small at that field region. Whether this suppression of the Nernst effect is simply from the suppression of the FISDW conduction channel or it is related to the sign reversal of the Hall effect is not yet clear.

In summary, we find the resonant Nernst effect is a general feature of several members of the Bechgaard salts in both their metallic and FISDW phases, where comparisons can be made. If the vortex model is applicable, p-wave Cooper pairing in the FISDW phase is implied. Since there is precedence for large Nernst signals in materials with density wave or non-Fermi liquid properties (including where the Hall sign is changing), the role of electron and hole carriers may be significant, or essential to the Nernst behavior observed. Although at present only the vortex model can explain the resonant Nernst sign reversal, the peaks in the torque measurements previously measured at the magic angles [12] do not follow from this model. Further work on $ClO_4$ under pressure would be useful to explore the anion ordering and FISDW threshold field parameters.

Acknowledgement
This work is supported by NSF 02-03532, the NHMFL is supported by the National Science Foundation and the State of Florida. The work at Ewha University is supported by KOSEF R01-2003-000-10470-0(2004).

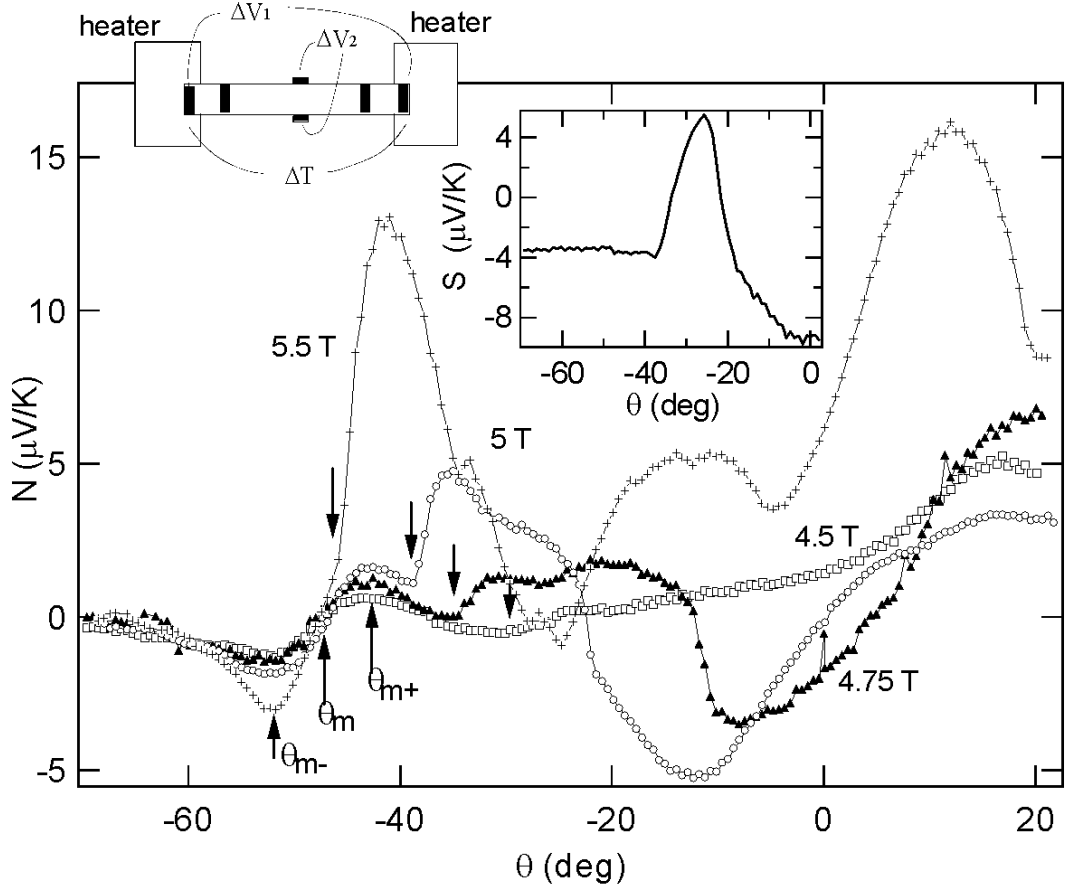

Fig. 1. Angle dependence of the Nernst signal at different fields at 0.4 K. The down arrows indicate the threshold angle ($\theta_{th}$) for the FISDW transition. The up arrows indicate angles where the Nernst signal shows positive peaks ($\theta_{m+}$) or negative dips ($\theta_{m-}$) in the metallic state around the magic angle for $p/q=1$ ($\theta_m$). Left inset: the contact configuration for the TEP and Nernst effect measurement. Right inset: the angle-dependent TEP (S) at 0.4 K and 5 T.

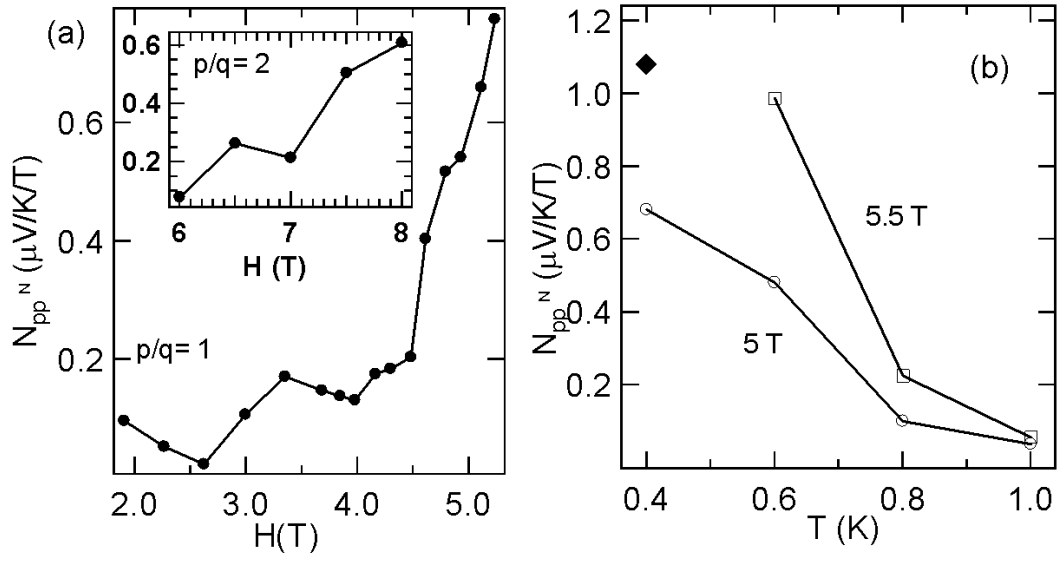

Fig. 2. The normalized peak-to-peak Nernst signal $(N(\theta_{m+})-N(\theta_{m-}))/H$ in the metallic state. a) Field dependence at 0.4 K for $p/q$ = 1 and 2 (inset). b) Temperature dependence at 5 T and 5.5 T. The data point at 0.4K and 5.5T (♦) is for the negative peak at $\theta = \theta_{m-}^{(1)}$ (since $\theta_{m+}^{(1)}$ was in the FISDW state).

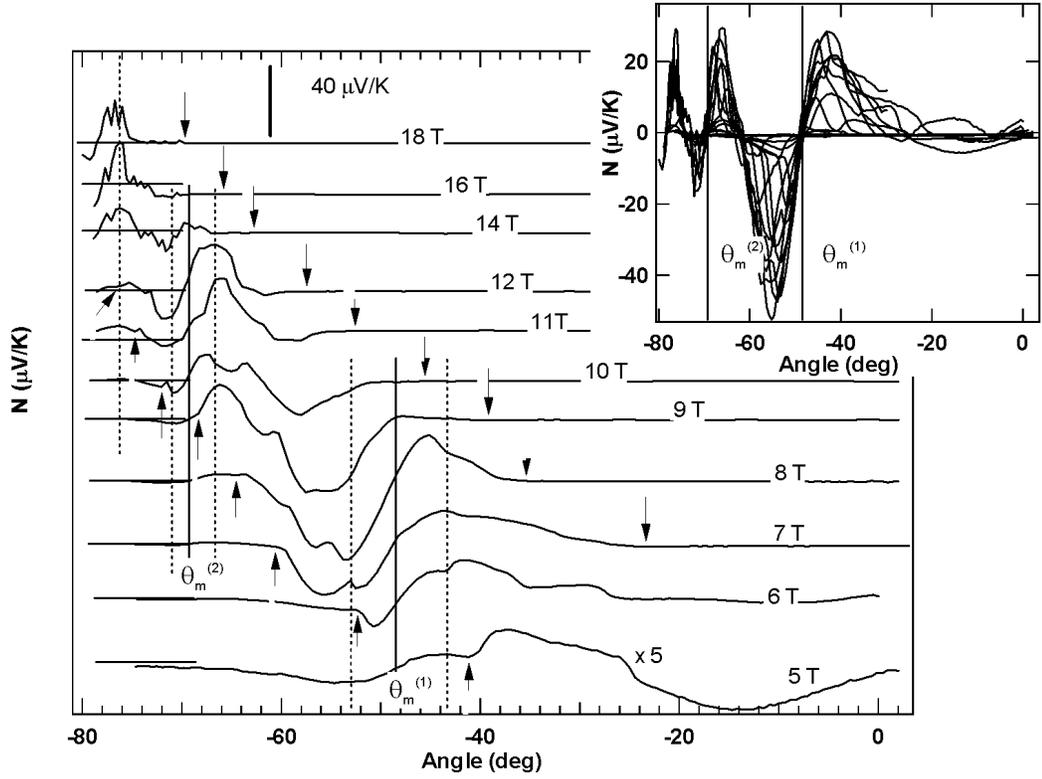

Fig. 3. The angle dependence of the Nernst effect for representative fields, offset for clarity, between 5 and 18 T at 0.4 K. The solid lines indicate the magic angles ($\theta_m^{(1)}$ for $p/q = 1$ and $\theta_m^{(2)}$ for $p/q = 2$) obtained from the ADMR measurements. The dotted lines indicate the angles where the Nernst signal shows a dip or peak in the metallic state. Up arrows: FISDW threshold angle $\theta_{th}$ where $H_\perp \sim 3.9$ T; down arrows: critical angle above which the Nernst effect is angle independent for $H_\perp \sim 6.5$ to 6.9 T. Inset: Nernst data for all fields between 5 and 18 T without offset.

S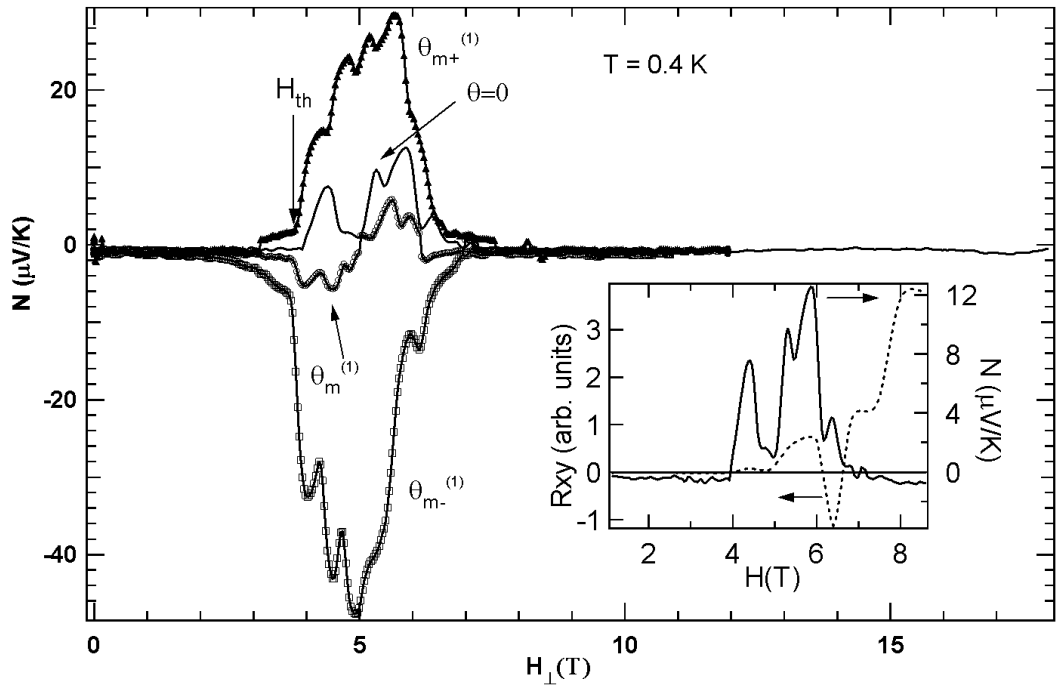

Fig. 4. The Nernst signal at 0.4 K as a function of $H_\perp$ (=$H \cos\theta$) obtained from field sweeps at fixed angles. The inset shows the Hall resistance and Nernst effect on the same sample for $\theta = 0$. The Ribault step [22] is the negative Hall signature near 6.5 T.

12